# A Catalogue of M51 type Galaxy Associations

Ari Jokimäki
Espoo, Finland (*email:  arijmaki@yahoo.com*)

Harley "Skip" Orr
Solon, ME, USA (*email:  skipandbeth@tds.net*)

David G. Russell
Owego Free Academy, Owego, NY USA (*email: russeld1@OACSD.org*)

**Abstract**

A catalog of 232 apparently interacting galaxy pairs of the M51 class is presented. Catalog members were identified from visual inspection of multi-band images in the IRSA archive.  The major findings in the compilation of this catalog are (1) A surprisingly low number of the main galaxies in M51 systems are early type spirals and barred spirals. (2) Over 70% of the main galaxies in M51 systems are 2-armed spirals.  (3) Some systems that were classified as M51 types in previous studies are not M51 types as defined in this catalog. There were a number of systems previously classified as M51 systems for which the "companion" is identified as an HII region within the main galaxy or foreground star within the Milky Way.  (4)  It was found that only 18% of the M51 type companions have redshift measurements in the literature.   There is a significant need for spectroscopic study of the companions in order to improve the value of the catalog as a sample for studying the effects of M51 type interaction on galaxy dynamics, morphology, and star formation.   Further spectroscopy will also help constrain the statistics of possible chance projections between foreground and background galaxies in the catalog.    The catalog also contains over 430 additional systems which are classified as "possible M51" systems.    The reasons for classifying certain systems as possible M51 systems are discussed.

## 1. Introduction

Interacting galaxies are interesting laboratories for the study of galaxy dynamics and evolutionary processes and have prompted the creation of a number of specialized catalogs including three catalogs of Vorontsov-Velyaminov (Vorontsov-Velyaminov, 1959, 1977, Vorontsov-Velyaminov *et al.* 2001 – hereafter VVI, VVII, and VVIII respectively), the "Atlas of Peculiar Galaxies" (Arp, 1966 – hereafter ARP), and the "Catalogue of Southern Peculiar Galaxies and Associations" (Arp & Madore, 1987 – hereafter AM).  Among the most intriguing examples of interaction are those represented by the archetype M51 – a large spiral galaxy with a companion at or near the end of a spiral arm. The VVI-III, ARP, and AM catalogs all include M51 type galaxies as a distinct class of associations and there have been a number of papers written about this class of objects, reporting studies of morphology, kinematics, photometry, and spectroscopy (Vorontsov-Vel'Yaminov  1976; Laurikainen, Salo,& Aparicio 1998; Klimanov & Reshetnikov 2001; Laurikainen & Salo 2000, 2001; Reshetnikov & Klimanov 2003; Rampazzo et al 2005; Fuentes-Carrera et al 2007).

While some M51 type galaxy pairs may be chance alignments of foreground and background objects projected near each other, there are often visible tidal disturbances, asymmetries in the spiral arms, and regions of accelerated star formation that provide additional evidence for interaction in these systems. As such, these systems may offer valuable information about the gravitational dynamics of galactic systems (eg. Fuentes-Carrera et al 2007) and perhaps address open questions about the effects of interaction on galactic evolution and as a triggering mechanism for star formation (eg. Hancock et al 2007; Smith et al 2007). Currently, a researcher interested in examining this class of objects must refer to any number of catalogs of interacting galaxies or any of the larger heterogeneous catalogs to select appropriate candidates for study. Motivated by the potential research value of a more comprehensive catalog of such associations, we have compiled a catalog of 232 M51 type systems.

This paper is organized as follows: Section 2 is a description of the criteria for adopting galaxies into the M51 catalog and the procedures utilized for determining whether or not candidates should be included. Section 3 is the description of the final catalog. Section 4 is the discussion, including the application of some of the statistical tests employed in Klimanov & Reshetnikov (2001) (hereafter K&R) to our sample. A brief conclusion is provided in section 5.

## 2. Selection Criteria and Methodology

### 2.1 M51 candidate samples

We have compiled a new catalog of M51 type galaxy associations by inspecting the morphology of galaxies in the catalogs of Vorontsov-Velyaminov (1959, 1977; Vorontsov-Velyaminov et al. 2001), Arp (1966), Arp&Madore (1987). The VV, ARP, and AM catalogs contain numerous categories, not all of which were considered as candidates in this study. The primary candidate sample included objects classified as M51 type objects in the VV, ARP, and AM catalogs. Specifically the primary sample included all objects classified as M, MM, and MMM in the VVIII (which includes also the M51 type systems from VVI and VVII); ARP 37-101; and AM category 9. Note that ARP catalog numbers 92-101 have not been always considered as M51 types, but we included them to our primary sample because they are very similar to M51 types. The secondary candidate sample included ARP 1-36 and 102-338 and AM categories 3b, 11, and 17.

Additional candidates were considered from the New General Catalog (Dreyer 1888) and a sample of 318 ScI galaxies with $K_S$-band Tully Fisher distances from the sample of Russell (2007) and a number of cases were serendipitous objects in the same field as candidate galaxies.

### 2.2 M51 classification criteria

The selection criteria were chosen to produce a catalog that contains galaxy associations consistent with the archetype of the class – M51, a spiral galaxy with a much smaller companion at or near the end of an arm. Galaxies included in the catalog met the following morphological criteria:

1)      The main galaxy must be a spiral.
2)      The angular diameter of the companion galaxy must be less than 50% of the angular diameter of the main galaxy.
3)      An arm from the main galaxy apparently connects and ends at the main body of the companion galaxy (M51bc, designated as such to indicate "bridge connected"). Note that we also accepted cases where bulk of the arm's luminosity ends at the companion even if a faint portion of the arm continues beyond the companion. Catalog object #19 (AM 0108-383) is an example of such a case.
4)      As an alternative to 3, an arm from the main galaxy points to but does not connect with the companion galaxy (M51bnc, designated to indicate the "bridge not connected").   M51bnc systems do not show a definite connection in visible wavelengths but the morphology often suggests a possible interaction between the two galaxies.   In practice M51bnc galaxies comprise only 16% of the galaxies in the catalog.

    Typical M51bc and M51bnc are shown in Figures 1a (catalog object #229 – ARP 86) and 1b (catalog object #97 – ARP 82).   Note that catalog object #97 is an example of an M51bnc with arm structure suggesting a possible tidal interaction between the main galaxy and the companion.    Figure 1c is catalog object #126 (NGC 3423) which is an example of M51bnc that lacks any apparent evidence for tidal interaction between the arm and companion in optical images.    The surface brightness of an arm is an important factor when differences between bc and bnc types are considered. It is probable that some of the bnc cases will turn out to be bc cases with deeper imaging or imaging in other wavelengths.    For example Sancisi *et al.* (2008) note that some galaxies possess clean optical images, but evidence for interaction in 21 cm HI emission.
    Many of the M51 candidates examined in the compilation of this catalog were previously cataloged as M51 type associations in the VV, ARP, or AM catalogs.    However, these systems were not automatically accepted as M51 type objects for this catalog until it was determined that they in fact met the criteria listed above.    Every M51 candidate was examined by visually inspecting its appearance in the multi-band DSS imagery from the NASA/IPAC Infrared Space Archive (IRSA).   Each candidate was observed by at least two members of the team (Jokimäki and Orr). For the sake of consistency, one of us examined each candidate in detail in multiple bands on at least one occasion using the IRSA "reproject" option.  In cases of disagreement over whether a galaxy association met the morphological guidelines**,** all three members of the team examined the association with the IRSA reproject option.    During the evaluation of these associations, we often compared the DSS images with those from other on-line sources, including the Sloan Digital Sky Survey, Hubble Space Telescope images, and other sources.
    Candidate pairs that unambiguously met the morphological criteria are part of the catalog of M51 associations presented in Table 1.    In many cases M51 candidates could not be classified as or rejected as M51 type systems with certainty.    In order to preserve these systems as subjects for future research, they were classified as "possible M51 associations" and cataloged in Table 2. See section 3 for descriptions of Table 1 and Table 2.    Possible M51 systems are pairs that fall into one of the categories summarized in Table 3.

Table 3. Possible M51 categories.

| Category | N cases | Explanation |
|----------|---------|-------------|
| 1 | 124 | The existence of an apparent physical association was uncertain or ambiguous with the IRSA reproject images. In many cases these systems are objects of small angular diameter that require higher resolution imagery. |
| 2 | 126 | The companion candidate was possibly a large HII region. Companions were flagged as potential HII regions when they were found to fade rapidly as images were examined from the B-band to the R-band to the IR-band. |
| 3 | 82 | The companion was stellar in appearance but could not be certainly identified as a star. |
| 4 | 43 | The main galaxy was not clearly identified as a spiral. |
| 5 | 44 | Borderline cases for which a clear determination could not be made. There were various reasons for this such as the companion's association with the end of the spiral arm being uncertain, but not sufficiently so as to warrant a clear rejection. |
| 6 | 16 | The system is oriented almost edge-on making its morphology and/or possible interaction difficult to determine. |

We believe there is potential research value in obtaining higher resolution multi-band imagery and spectroscopy not only for the objects in the M51 catalog (Table 1), but also for the possible M51 candidates (Table 2). For example we note the recent discovery by Madore et al (2007) that the object NGC 6908 - which was long thought to simply be a "surface brightness enhancement" in a spiral arm of NGC 6907 - is in fact a lenticular galaxy.

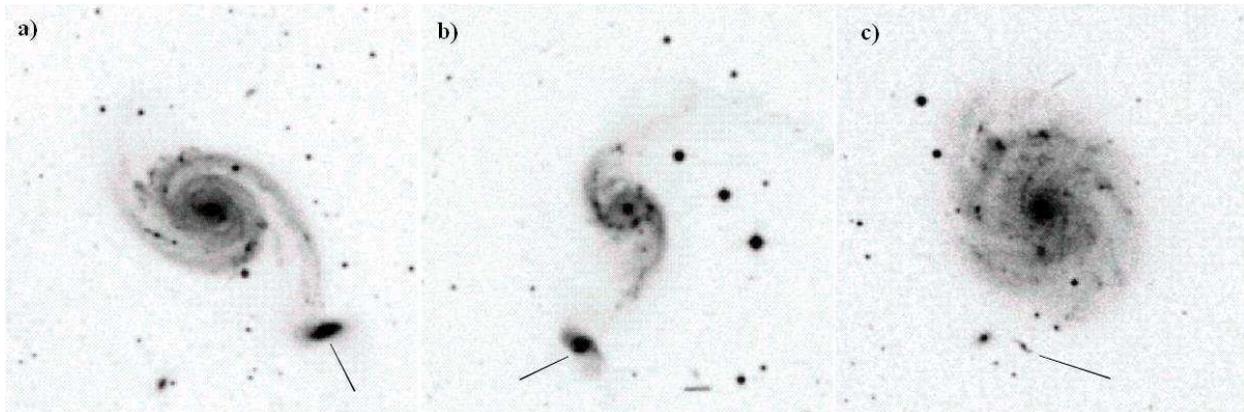

Figure 1a - M51bc          1b - M51bnc          1c - M51bnc

## 2.3 Rejected Candidates

Galaxy associations from the candidate galaxy lists were rejected from the M51 type catalog for a wide range of reasons. The following is the list of reasons candidates were rejected from the final M51 catalog. The candidates rejected for each of these reasons are listed in Table 4. Note that rejected systems may have more than one reason for rejection, but only the primary rejection reason is indicated in Table 4.

(1) The companion is not located at the end of the arm. In such cases the arm either extends past the companion and appears unaffected by interaction, or misses the potential companion. In a few extreme examples, there is no arm to the companion at all. In a couple of cases there are two overlapping galaxies but no association with the arms.

(2) The candidate companion is not a galaxy but instead is a foreground star within the Milky Way, HII region within the main galaxy, brightening in the arm of the main galaxy, or tidal feature. Note that these cases are such that we can be sure about the companion's status. There are similar cases in Table 2 for which we cannot be sure about the companion's status.

(3) The main galaxy is not a spiral galaxy.

(4) There is no obvious companion galaxy.

(5) The system was not found.

(6) The angular diameter of the companion galaxy is not less than 50% of the angular diameter of the main galaxy.

(7) Connecting bridge is not a spiral arm.

Table 4. Rejected systems identified by rejection reason.

| Reason for rejection | Systems rejected |
|---|---|
| (1) The companion is not located at the end of the arm. | ARP 043, ARP 044, ARP 054, ARP 064, ARP 065, ARP 068, ARP 069, ARP 071, ARP 072, ARP 073, ARP 076, ARP 077, ARP 078, ARP 083, ARP 084, ARP 095, AM 0017-454, AM 0020-695, AM 0037-305, AM 0215-331, AM 0459-340, AM 0500-590, AM 0520-222, AM 0536-502, AM 0639-582, AM 0654-281, AM 0729-660, AM 1237-364, AM 1255-241, AM 1259-340, AM 1353-272, AM 1357-253, AM 1401-324, AM 1416-262, AM 2041-530, AM 2046-680, AM 2057-650, AM 2100-381, AM 2259-243, AM 2302-515, AM 2319-595, AM 2341-252, AM 2343-633, AM 2357-682, VV 003, VV 006, VV 017, VV 019, VV 369, VV 378, VV 381, VV 381, VV 400, VV 415, VV 429, VV 431, VV 438, VV 440, VV 448, VV 458, VV 465, VV 473, VV 484, VV 679, VV 986, VV 1142, VV 1242, VV 1253, VV 1292, VV 1345, VV 1480, VV 1640, VV 1649, VV 1802, VV 1870. |
| (2) The candidate companion is not a galaxy | VV 008, VV 029, VV 436, VV 456, VV 465, VV 477. |
| (3) The main galaxy is not a spiral galaxy. | AM 0244-612, AM 0456-264, VV 089, VV 382. |

| (4) There is no obvious companion galaxy. | ARP 060, ARP 099, ARP 101, AM 0305-824, AM 0507-512, AM 0827-845, AM 1000-254, AM 1255-503, AM 1311-365, AM 1604-324, AM 2052-301, AM 2339-583, VV 056, VV 366, VV 367, VV 373, VV 376, VV 377, VV 383, VV 384, VV 387, VV 389, VV 391, VV 394, VV 405, VV 416, VV 419, VV 430, VV 432, VV 433, VV 441, VV 467, VV 875, VV 974, VV 1067, VV 1228, VV 1251, VV 1277, VV 1316, VV 1354, VV 1417, VV 1424, VV 1478, VV 1486, VV 1490, VV 1494, VV 1497, VV 1508, VV 1533, VV 1541, VV 1551, VV 1580, VV 1681, VV 1698, VV 1747, VV 1766, VV 1784, VV 1815, VV 1842, VV 1871, VV 1949, VV 1968, VV 1980, VV 1984. |
|---|---|
| (5) The system was not found | AM 1332-290 and VV 437 |
| (6) The angular diameter of the companion galaxy is not less than 50% of the angular diameter of the main galaxy | ARP 040, ARP 045, ARP 087, ARP 091, ARP 093, ARP 096, ARP 098, AM 0013-562, AM 0044-521, AM 0120-325, AM 0158-345, AM 0207-360, AM 0208-223, AM 0215-320, AM 0218-321, AM 0244-530, AM 0322-670, AM 0324-362, AM 0324-524, AM 0327-285, AM 0412-474, AM 0417-754, AM 0422-275, AM 0448-622, AM 0523-400, AM 0558-600, AM 0836-271, AM 0849-272, AM 1017-271, AM 1823-512, AM 2025-370, AM 2218-840, AM 2219-255, AM 2219-571, AM 2311-472, AM 2327-422, AM 2341-252, VV 386, VV 410, VV 414, VV 458. |
| (7) Connecting bridge is not a spiral arm | ARP 046, ARP 055, ARP 081, ARP 089, ARP 090, ARP 097, AM 0125-850, AM 0523-244, AM 0536-502, AM 0606-770, AM 0601-320, AM 0623-584, AM 1108-270, AM 1108-300, AM 1344-323, AM 2052-221, AM 2103-474, AM 2105-332, AM 2144-360, AM 2219-432, AM 2233-613, AM 2245-401, AM 2256-480, VV 020, VV 022, VV 365, VV 371, VV 374, VV 814, VV 927, VV 1255, VV 1270, VV 1333, VV 1464, VV 1562, VV 1837, VV 1843. |

## 3. The Catalog of M51 Associations.

The catalog of M51 associations is presented in Table 1.  The columns in Table 1 are as follows:

Column 1: Catalog number.  For main galaxies with more than one M51 type companion there will be more than one entry for the main galaxy in the table.
Column 2: Designation of the main galaxy.
Column 3: Morphological type of the main galaxy.
Column 4: Heliocentric radial velocity of the main galaxy (km s$^{-1}$).
Column 5: The major diameter of the main galaxy (arcmin). Major diameters have been taken from HyperLeda, and from NED if not available in HyperLeda.
Column 6: Designation of the companion galaxy.

Column 7: Heliocentric radial velocity of the companion galaxy (km s$^{-1}$).
Column 8: The major diameter of the companion galaxy (arcmin). Major diameters have been taken from HyperLeda, and from NED if not available in HyperLeda.
Column 9. Angular separation between the main galaxy and the companion (arcmin).
Column 10. Position Angle of the companion in relation to the main galaxy (degrees). Positions are given as degrees from the north of the main galaxy (0 deg is N, 90 deg is E, 180 deg is S, and 270 deg is W).
Column 11: Type of the M51 association (bc/bnc).
Column 12: Designations of the system in different catalogs.

Possible M51 candidates are presented in Table 2.   The columns for table 2 are as follows:

Column 1: Catalog number.   For galaxies with more than one possible M51 type companion there will be more than one entry for the main galaxy in the table.
Column 2: Designation of the main galaxy.
Column 3: Designation of the companion galaxy.
Column 4. Angular separation between the main galaxy and the companion (arcmin).
Column 5. Position Angle of the companion in relation to the main galaxy (degrees). Positions are given as degrees from the north of the main galaxy (0 deg is N, 90 deg is E, 180 deg is S, and 270 deg is W).
Column 6:  Reason for the uncertainty of the system as an M51 system.   Reasons are numbered as given in Table 3.
Column 7:  Designation of the system in different catalogs.

Please note that these tables are truncated to facilitate publication.  The full tabular data for both Table 1 (M51) and Table 2 (possible M51) galaxy associations is available at www.jorcat.com. Annotated images of all Table 1 and Table 2 cases are also available at the same web site.

## 4.  Sample Statistics

### 4.1  Candidate Sample Classification Statistics

The initial candidate sample included a total of 867 systems with at least 1099 candidate companion galaxies inspected. Table 5 is a presentation of the final classification statistics for the candidate systems and galaxies. Column 1 of Table 5 is the subset in question. Columns 2 and 3 are the numbers of systems and cases (respectively) in whole sample. "Case" here refers to an individual candidate object near the main galaxy. A main galaxy and all the candidate objects near it constitute a "system". Column 4 is the number of systems that have M51 cases and column 5 is the number of M51 cases (Table 1 cases). Column 6 is the number of systems that have possible M51 cases and column 7 is the number of possible M51 cases (Table 2 cases). Column 8 is the number of systems that have rejected cases and column 9 is the number of rejected cases. It should be noted that the rejected cases in column 9 include candidates that were promising M51 candidates requiring close inspection before rejecting. In some cases there were companion galaxies in the field of the main galaxy but not near the end of a spiral arm. Those companions were rejected right away without closer inspection, and they are also not included to given numbers.

Table 5: Case statistics of the M51 catalog.

| Subset | Whole sample | | M51 types (T1) | | Possible (T2) | | Rejected | |
|---|---|---|---|---|---|---|---|---|
| | Systems | Cases | Systems | Cases | Systems | Cases | Systems | Cases |
| Whole sample | 867 | 1099 | 208 | 232 | 347 | 435 | 414 | 432 |
| ARP | 38 | 49 | 12 | 14 | 8 | 11 | 23 | 24 |
| AM | 168 | 217 | 55 | 62 | 64 | 74 | 78 | 81 |
| VV | 222 | 295 | 47 | 50 | 101 | 135 | 104 | 110 |
| ARP & VV | 26 | 37 | 20 | 23 | 6 | 6 | 8 | 8 |
| ARP & AM | 1 | 1 | | | | | 1 | 1 |
| AM & VV | 3 | 3 | 2 | 2 | 1 | 1 | 0 | 0 |
| ARP secondary | 20 | 28 | 1 | 1 | 7 | 10 | 14 | 15 |
| AM secondary | 26 | 32 | 8 | 8 | 7 | 8 | 13 | 16 |
| VV secondary | 13 | 15 | | | 3 | 3 | 10 | 12 |
| Additional NGC * | 268 | 327 | 53 | 60 | 106 | 132 | 133 | 135 |
| RUSSELL | 63 | 75 | 6 | 6 | 36 | 46 | 23 | 23 |
| EXTRA | 19 | 22 | 4 | 6 | 8 | 9 | 7 | 7 |
| | | | | | | | | |
| Previous M51 types | 458 | 602 | 136 | 151 | 180 | 227 | 214** | 224 |
| New | 409 | 497 | 72 | 81 | 166 | 208 | 200 | 208 |
| | | | | | | | | |
| NGC total | 389 | 492 | 81 | 131 | 150 | 231 | 199 | 233 |

*There are many NGC systems in ARP, AM, and VV catalogs, these values are additional to them. The last row of the table gives the total NGC values.*
**This number contains only 188 systems for which all obvious M51 type candidates were rejected. The additional 26 systems are such that they have at least one obvious M51 type candidate classified either M51 type or possible M51 type, even if they have some candidate companions rejected.*

The following subsets are given in Table 5: The first row is the total numbers for the whole sample. Then the numbers for different catalogs of origin are given. These have been explained in section 2. Additionally, Table 5 has rows for "ARP & VV", "ARP & AM", and "AM & VV". These subsets are included because some systems are present in two catalogs of origin. We note that only the ARP and VV catalogs overlap significantly for M51 type systems (26 candidate systems in both catalogs). Interestingly, these 26 systems had the highest percentage of acceptance into this M51 catalog with 20/26 candidates (76%) yielding M51 companions compared with only 32% for the ARP only, 33% for the AM only, and 21% for the VV only systems. Next we have included a comparison of previously classified as M51 types, and the new M51 type systems proposed by us. Our selection criteria resulted in 72 new M51 type systems. Among the systems previously classified as M51 types (458 systems), 30 % (136 systems) were classified as M51 type systems. From Table 5 it can be calculated that of the original 65 M51 systems identified in the ARP catalog, we accepted 32 (49 %) as M51 types. The VV catalogs have a combined 251 objects identified as M51 systems, of which we accepted 69 (27 %). The AM catalog has 172 M51 type systems, of which we accepted 57 (33 %). We think that the difference in these percentages between ARP catalog and VV and AM catalogs is easily explained by the fact that ARP catalog contains mainly bright nearby systems, whereas VV and AM catalogs contain more faint systems which are more difficult to classify correctly. We note that the mean B-band magnitudes of the ARP, AM, and VV catalogs are 13.48, 15.02, and 14.08 respectively which supports this explanation.

The last row gives the total number of M51 type systems found among NGC galaxies. With that number we can give a rough estimate of the total fraction of M51 type systems among all galaxies. There are 6038 galaxies in NGC catalog and 3711 of these are spirals (based on the information found on The NGC/IC Project website). We have classified 81 systems from the NGC catalog as M51 types which gives an estimate that 1.3 % of all galaxies in the NGC catalog and 2.2% of all spirals in the NGC catalog are of the M51 type.

### 4.2 Morphology statistics

Table 6 provides the morphology distribution of the main galaxies in the catalog. K&R noted that early type spirals (Sa or Sab) are almost absent in their sample - being represented by only 3% of their 32 sample G M51 galaxies. We find that 29 of the 208 main galaxies (13.9%) that have been classified as M51 systems are of type Sa/Sab. The difference is not surprising given that the sample in this paper is over six times larger. However, it is evident in our sample that – as suggested from the K&R sample - M51 systems are dominated by later type Sb/Sbc/Sc systems. We find that 72.6% of our M51 systems are of types Sb, Sbc, and Sc. For possible M51 systems we find that 8.6% are types Sab/Sb and 71.6% are of types Sb/Sbc and Sc.

Table 6. Distribution of main galaxy types in M51 catalog.

|  | M51 galaxies (Table 1) | | Possible M51 galaxies (Table 2) | |
| --- | --- | --- | --- | --- |
|  | number | percentage | number | percentage |
| Total | 208 | 100 | 349 | 100 |
| Sa | 17 | 8 ±2 | 20 | 6 ±2 |
| Sab | 12 | 6 ±2 | 10 | 3 ±1 |
| Sa + Sab | 29 | 14 ±3 | 30 | 9 ±2 |
| Sb | 54 | 26 ±3 | 68 | 19 ±3 |
| Sbc | 46 | 22 ±3 | 53 | 15 ±2 |
| Sc | 51 | 25 ±3 | 129 | 37 ±3 |
| Scd | 0 | 0 | 6 | 2 ±1 |
| Sd | 0 | 0 | 5 | 1 ±1 |
| Others | 28 | 13 ±3 | 58 | 17 ±2 |

As another comparison and following Elmegreen, Elmegreen & Bellin (1990-hereafter EEB90) procedures, we calculated the early type galaxy fractions for our M51 and possible M51 samples. Our M51 sample has 83 Sa-b type main galaxies and 97 Sbc-cd type main galaxies. Therefore the early type galaxy fraction is 0.46 ±0.04 in our M51 sample when we include Sb galaxies in the early type category as was done by EEB90. Our possible M51 sample has 98 Sa-b type main galaxies and 188 Sbc-cd type main galaxies. That makes early type galaxy fraction 0.34 ±0.03 in our M51 sample. These numbers are low compared to EEB90 numbers for binary galaxies, but similar to their fractions for field galaxies although slightly smaller in the possible M51 sample. K&R noted that the scarcity of early spirals could be a selection effect from selecting galaxies with distinct spiral structure. That is also a possibility with our sample.

The group "others" in Table 6 primarily includes galaxies which had uncertain or unknown types listed in HyperLeda and/or NED databases, but there are also a few that are listed as irregulars (galaxy types in our sample are given in Tables 1 and 2). The total numbers of 208 for Table 1 galaxies and 349 for Table 2 galaxies represent the sample size of the main galaxies. There are several main galaxies in Tables 1 and 2 that have more than one companion listed in each table.

Table 1 has 80 barred galaxies and 22 galaxies with unknown bar situation out of a total of 208 main galaxies, which makes the bar fraction (EEB90) 0.30 ±0.03. This is less than what EEB90 report for binary systems, and once again similar to their fractions of field galaxies.

It is interesting to note that with regard to both the fraction of early type spirals and the fraction of barred spirals, the M51 and possible M51 samples in our catalog are smaller than the fractions reported for binary galaxies by EEB90 but similar to the fractions reported for field galaxies. This might seem surprising given that the M51 systems in this catalog are in many cases binary systems. One possible explanation for the smaller fraction of barred spirals in the M51 sample relative to the EEB90 binary sample is that the M51 sample contains many small companions which may have insufficient mass to trigger the formation of bars in the main galaxies. On the other hand, the EEB90 samples generally show much higher bar fractions in early type spirals than in Sbc/Sc spirals. Given, that the early type spiral fraction is somewhat lower in the M51 catalog, it might then be expected that the bar fraction should also be smaller. Finally, it is also worth noting that the binary systems in the EEB90 sample are not specifically of the M51 type and thus there is no a priori reason to expect the M51 sample to have the same morphology distribution as the EEB90 binary galaxy sample.

### 4.3 Arms of Main Galaxies

Arm counts in our sample are presented in Table 7. Similar to the findings of K&R, two-armed spirals are clearly most common in M51 type galaxies. The distribution of arm counts of our Table 1 galaxies overall is quite similar to distribution of K&R sample. However, while both the K&R sample and our Table 1 galaxies show that about 71% have 2 arms, in our Table 2 galaxies (possible M51 type associations) only 52 ±3 % of the main galaxies have 2 arms. Since a significant portion of the possible M51 associations may turn out to be rejected M51 associations with further study this result suggests that M51 type galaxies tend to have 2 arms.

Table 7. Arm counts of main galaxies in M51 catalog.

| arm count | M51 galaxies (Table 1) / % | Possible M51 galaxies (Table 2) / % |
|---|---|---|
| 1 | 12 ±3 | 12 ±2 |
| 2 | 71 ±4 | 52 ±3 |
| 3 | 13 ±3 | 21 ±3 |
| 4 | 5 ±2 | 9 ±2 |
| 5 or more | 1 ±1 | 2 ±1 |
| unclear | 0 | 4 ±2 |

In some cases it was difficult to determine whether the connecting bridge was a spiral arm, or just a tidal feature resembling a spiral arm. There were a few different types of cases exhibiting this. First, in some cases there were very loose spiral arms which made it difficult to assess whether the main galaxy was a spiral with very loose arms or an elliptical galaxy with tidal bridges and tails (e.g. Table 1 cases: #31, #37, #61, #129, #192, #231, Table 2 cases: #250, #511, and rejected cases: AM 0218-321, AM 0417-754, AM 2219-432). Second, two cases (#211 and rejected VV 1837) had small spiral arms and another set of tidal bridges or tails seemingly resembling much larger spiral arms. Third, there were some cases where the bridge was short, straight, and started tangentially from other side of the main galaxy, and usually there were some additional signs of interaction from the center of the main galaxy, and even from the other side as well (e.g. Table 1 cases: #49, #169, #223, and Table 2 cases: #259, #353, #364, #643).

## 4.4 Evaluation of the K&R M51 sample

A system-by-system comparison of the systems identified as M51 associations by K&R reveals that our selection criteria resulted in the rejection of about half of the systems K&R classified as M51 associations (Table 8) in their sample G. Specifically, of the 32 systems K&R classified as M51 systems, we accepted 17 systems as M51 systems, accepted 5 systems as possible M51 systems and rejected 8 systems from the M51 catalog. There were two systems (K&R #'s 5 and 11) for which the companion identified by K&R was rejected, but for which a smaller companion at the end of an arm was accepted to our possible M51 sample but flagged as a possible HII region.

Table 8  Rejected and possible M51 cases from the K&R sample

| K&R ID | Main Galaxy | Classification in present work | Comment |
|---|---|---|---|
| 5 | ARP 54 | Rejected | Arm does not point to the large companion |
| 5 | ARP 54 | Possible M51 | Small companion is a possible HII region |
| 9 | NGC 2864 | Possible M51 | Arm may miss companion |
| 11 | VV410 | Rejected | Arm misses large companion |
| 11 | VV410 | Possible M51 | Small companion is a possible HII region |
| 12 | UGC 6293 | Rejected | Arm not associated with companion |
| 13 | ARP 83 | Rejected | Arm does not lead to companion's main body |
| 14 | ARP 87 | Rejected | Companion is 77% of main galaxy diameter |
| 15 | ARP 62 | Possible M51 | Interaction likely, but may not be M51 type interaction |
| 16 | ARP 18 | Rejected | Companion is an HII region |
| 19 | VV431 | Rejected | Companion is a star |
| 22 | VV019 | Rejected | Arm does not lead to the companion's main body |
| 23 | NGC 5497 | Rejected | Arm misses companion |
| 26 | VV487 | Possible M51 | Arm extends slightly past companion |
| 27 | VV452 | Possible M51 | Arm may miss companion |
| 29 | VV 447 | Possible M51 | Companion appears to be a star |
| 32 | ARP 68 | Rejected | Arm misses companion |

## 4.5  Does our selection criteria improve the identification of true M51 systems?

As suggested by the referee, it is worth evaluating whether or not the procedure and selection criteria applied in the creation of this catalog improves the success rate of identifying truly interacting systems of the M51 type. At a minimum, we believe that the utilization of multi-band imagery from the DSS and other sources such as SDSS reduces the number of false M51 cases by allowing for identification of "companions" that are in fact only HII regions or foreground stars. For example, we note that the companion of object #16 from the K&R sample (Arp 18) is clearly an HII region in the multi-band images of the IRSA archive. This is readily evident as the companion significantly fades in the R and IR bands compared with the B-band image (Figure 2).

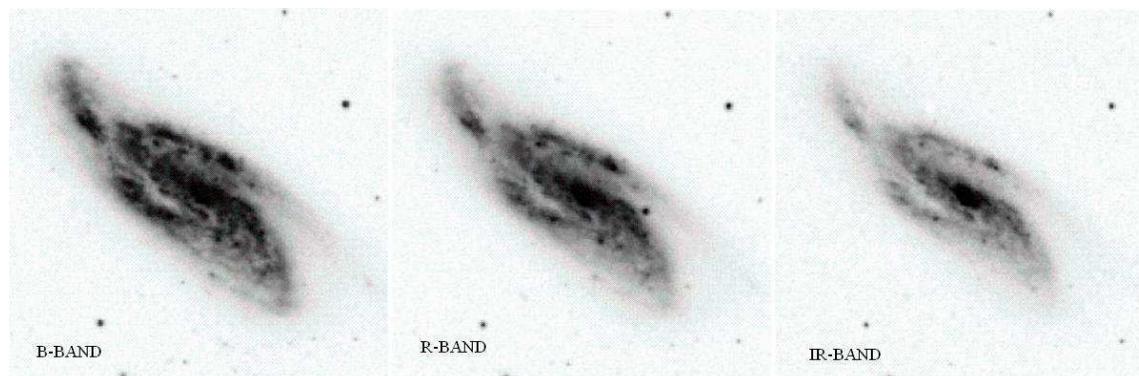

Figure 2 – B,R, and IR band images of Arp 18. The candidate companion fades strongly in the IR band and is a probable HII region.

As another example, the companion object for K&R object #19 appears to be a star as it shows negligible change in appearance in the B, R, and IR bands (Figure 3). In the SDSS catalog, the companion to object #19 is *SDSS J121802.24+063859*.0 and is identified as a star. While objects 16 and 19 were included in the K&R sample, the procedures adopted in this study allowed us to confirm that these objects are not galaxies and thus should be rejected from the M51 catalog.

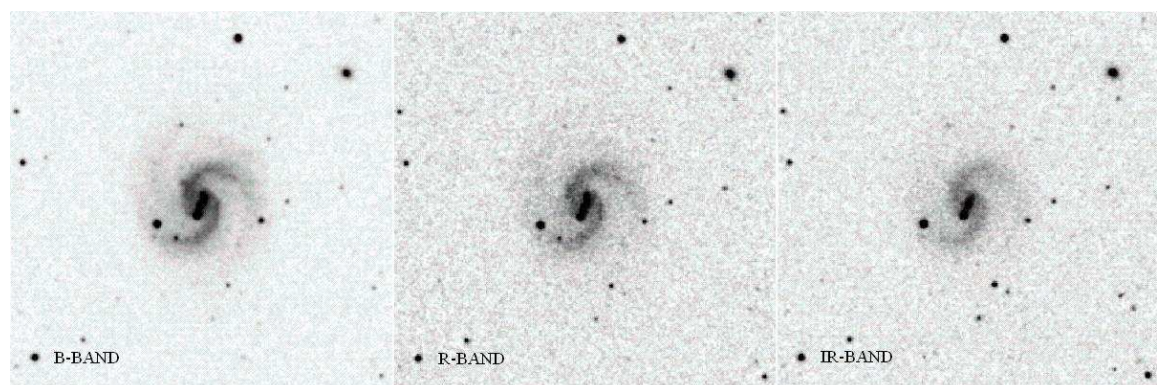

Figure 3 – B, R, and IR band images of K&R object #19. The companion is classified as a star in the SDSS.

As a further illustration of the importance of the multi-band imagery, the curious reader is encouraged to examine the IRSA B,R, and IR band DSS2 images of NGC 6907 (see Madore et al 2007). Whereas the identity of NGC 6908 as a galaxy is obscured in the B-band and R-band images, the obscuring HII regions have faded in the IR band image allowing the nature of NGC 6908 as a lenticular galaxy to be revealed.

Seven K&R galaxies are rejected, and three K&R galaxies are listed as "possible" M51 associations because the companion does not appear at the end of a spiral arm (Table 8). While this particular reason for rejection may eliminate some candidates that others might judge to be M51 type systems we note that one goal of this catalog was to identify associations for which there is a possible interaction between the companion and the *end* of the spiral arm. In the case of the nearly face on object #32 (ARP 68) from the K&R catalog, it is apparent that the arm near the companion straightens long before the near contact point with the companion and extends slightly past the companion. Interaction between the main galaxy and the companion is not ruled out in this instance. However, there is no evidence that compels the conclusion that ARP 68 is an M51 type interaction specifically and thus the system was rejected from our catalog. We believe that the careful consideration of each system by two or more evaluators has helped to reduce the number of systems such as ARP 68 that might previously have been classified as M51 systems, but for which M51 type interaction is not evident in multi-band images. Table 4 may be consulted for rejected candidates that may be useful as candidates for study of non-M51 type galaxy interactions.

When the M51 samples of ARP, AM, and VV are considered, there are many similar cases as K&R cases 16, 19, and 32 discussed above that were rejected (section 2.3). In the rejection category 1, ARP 71 turns out to be a case of two chance projected, slightly overlapping spiral galaxies in the SDSS images. Among category 2 cases is VV 8, which seems to have a companion candidate at the end of an arm in VVI catalog and ARP atlas images. Even in DSS images it looks like possible companion, but as the companion seems to fade away in IR band, while being strong in B band, we start to think of it as an HII region. For this case there is an even better image available at National Optical Astronomy Observatory (NOAO) web site that clearly reveals the region in question is not a companion galaxy. Category 3 holds VV 382, which is quite clearly an elliptical galaxy, rather than M51 type spiral, in SDSS images.

## 4.6 Chance projections

Another issue is the possibility of chance projections of foreground and background objects. With the extreme incompleteness of the companion redshifts in the sample, it is difficult to properly assess this potential problem. For this reason, we have taken an approach that compares the probability for the companion galaxies' appearance in apparent interaction with their main galaxies to the distribution of background galaxies in general.

The probability of chance projection of background galaxies is calculated simply by dividing the area around main galaxy in which an object would be considered as M51 type companion by the area in background where on average one background galaxy is found. We calculated the chance projection probability in an average case within our M51 sample. We used rather simply and conservatively a ring around main galaxy as the potential M51 companion area (see Figure 4 for illustration). We divided our companion sample to two populations; to those that lie within main galaxy's semimajor axis, and to those that lie outside main galaxy's semimajor axis. We calculated the arithmetic mean deviation from main galaxy's semimajor axis for both populations ($r_{D1}$ = 0.21 arcmin for inner and $r_{D2}$ = 0.32 arcmin for outer population), and we calculated the arithmetic mean semi-major axis for all the main galaxies in our M51 sample (r = 1.4). We then used those to calculate the inner and outer radiuses for the M51 ring. The area of the ring ($A_{M51}$) is then the circle area of the inner radius subtracted from the circle area of the outer radius:

$$A_{M51} = \pi\left(r + r_{D2}\right)^2 - \pi\left(r - r_{D1}\right)^2 \qquad (1)$$

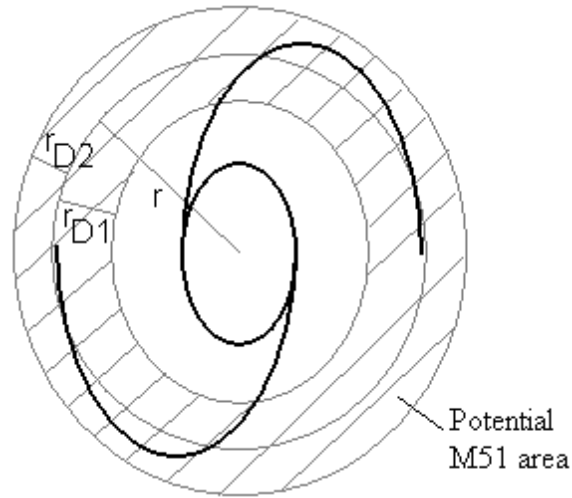

Figure 4 – Illustration of the potential M51 companion area around a main galaxy.

The calculated area from that is $A_{M51} = 1.9 \times 10^{-4}$ deg$^2$. We get the average area for one background galaxy ($A_B$) from number counts of galaxies. The average B band magnitude for the companions in our M51 sample (Table 1) is 16.45. Liske et al (2003) give Millenium Galaxy Catalogue galaxy count of 2.44 per square degree for B band magnitude 16.25 and 4.9 per square degree for 16.75 mag. As our average magnitude is quite close to halfway between these, we adopt the higher count for our calculation. Based on that galaxy count, the area of one background galaxy is $A_B = 1$ deg / (4.9 / deg) = 0.20 deg$^2$, and that gives us the probability of $A_{M51} / A_B = 0.001$ for chance projections in our M51 sample. It means that out of our 232 Table 1 cases, 0.2 are expected to be chance projections. The discordant redshift cases seem to be far more abundant than that, total of 17 of 232 (7 %), and for those that have measured redshifts 17 of 41 (41 %), have discordant redshifts. However, we have to remember that the probability calculated here is only a very strongly averaged estimate.

There are at least two things adding uncertainty to our calculated probability: 1) The Liske *et al*. (2003) galaxy count is for a 0.5 mag bin but our sample covers a much wider range which means that the actual chance projection probability would increase. Also, only 72 out of 232 cases in our M51 sample have B band magnitudes available, and it would make sense to assume that those companions that lack magnitude data are generally fainter than the companions that have magnitude data. That would make the average B band magnitude fainter than the currently calculated 16.45 and we would probably need to adopt a higher galaxy count which would increase the expected number of chance projections. 2) The area $A_{M51}$ we used is an inaccurate description of the real situation because we do not expect that M51 companions can lie anywhere around the main galaxy, instead they would have to lie near the ends of the arms of the main galaxy, as is determined by our selection criteria. That means that the actual area for M51 companions is smaller than the ring around main galaxy we used. How much smaller depends on arm counts of main galaxies, bc/bnc status and companion diameters of individual cases among other things. A smaller M51 companion area would therefore reduce the expected number of chance projections, so this uncertainty works in opposite direction of the first one.

Object #222 (ARP 56 system) has a companion with fainter than the average B band magnitude (17.84), the diameter of the companion is d = 0.37 arcmin, the main galaxy has two arms, and it is classified as M51bc in Table 1. The potential M51 area for this case is circle area defined by companion's semimajor axis (= d/2) multiplied by main galaxy's arm count n (because the companion could be at the end of either arm):

$$A_{M51} = n\pi(d/2)^2 \qquad (2)$$

From Liske *et al.* (2003) the galaxy count for 17.75 mag count is 16.77 per square degree. For this object #222 we then get a chance projection probability of 0.001. When we repeat this calculation for all our cases that have enough data available for this calculation, we get probability range from $1.9 \times 10^{-5}$ to $1.5 \times 10^{-3}$, and an arithmetic mean of $4.7 \times 10^{-4}$. This suggests that we might have been too conservative in our estimation of potential M51 area in our average calculation above. However, the bnc cases have larger potential M51 area (but difficult to estimate) than bc cases, so that would increase the mean probability in our sample.

Due to the nature of bnc cases, in which a spiral arm points to, but does not visibly connect to a companion galaxy, there is an enhanced possibility that the companion is a distant galaxy seen in projection, and is not in gravitational interaction with the main galaxy. The likelihood of such associations being chance projections is heightened when the spiral arm(s) involved are smooth and symmetrical, without evidence of enhanced star-formation, tidal disturbance, or other disruption (Figure 1c). We have elected to keep the bnc associations in the catalog with the understanding that bridges between galaxies are subject to gravitational dynamics and cannot be modeled as if they are permanent fixed structures. Another factor favoring inclusion of the bnc cases in the data set is the lack of spectroscopy on most of these companion galaxies. If a larger fraction of bnc cases (vs bc cases) are chance projections, it is expected that the incidence of discordant redshifts (>1000 k/sec) will be proportionally higher in the bnc cases. It is hoped that follow-up spectroscopy on these galaxies will shed light on this open question.

### 4.7  Companion positions

Following K&R, we calculated the angular separation to semimajor axis ratio which they called 1/S. The cumulative distribution of 1/S for Table 1 cases is shown in Figure 5. Most (63 %) of the companions are located within the semimajor axis defined area of main galaxies. A clear majority (81 %) of companions lie between 0.5 and 1.5 multiples of semimajor axis. We agree with K&R when they say: "Thus the satellites in M 51-type systems are situated near the boundary of the stellar disks of the main galaxies".

We calculated differentials of companion and main galaxy major axis position angles. Results are shown in Figure 6. We find that 41 % of the companion galaxies in the M51 sample lie within ±20 degrees in position angle of the major axis position angle, and only 10 % lie within ±20 degrees in position angle of the minor axis position angle. Corresponding numbers for the possible M51 sample are 46 % for major axis and 11 % for minor axis. This result agrees with previous studies (Yang *et al.* 2006).

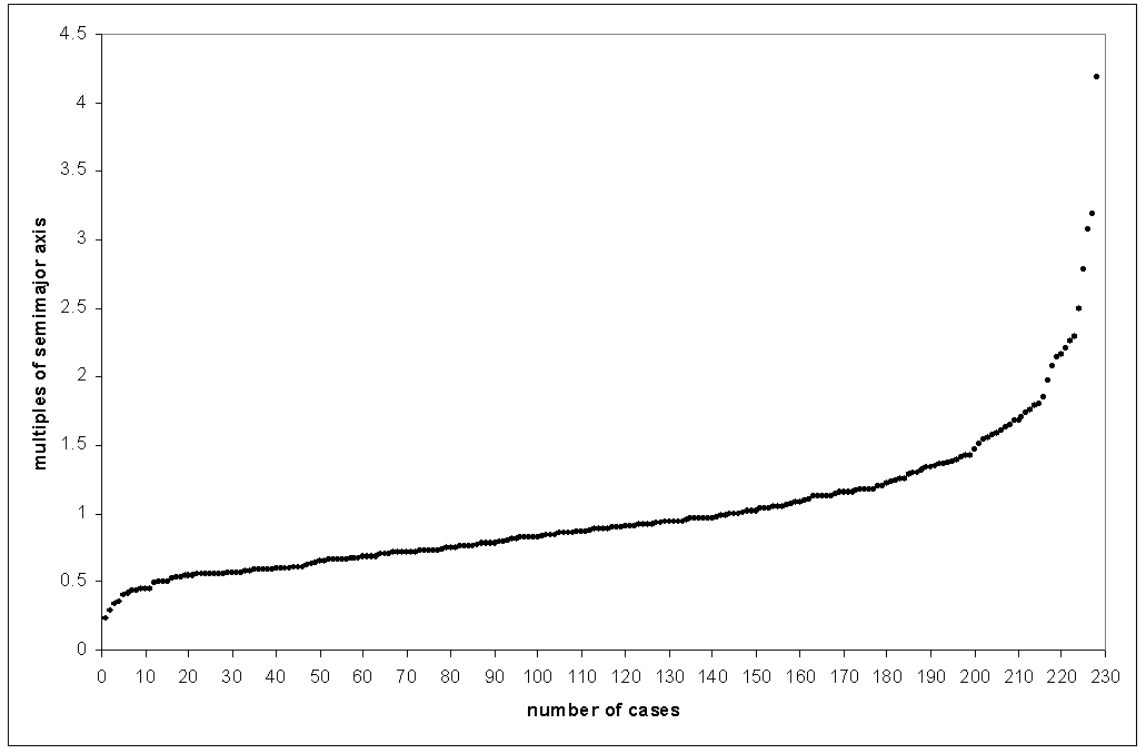

Figure 5 -  Cumulative distribution of the angular separation to semi-major axis ratio.

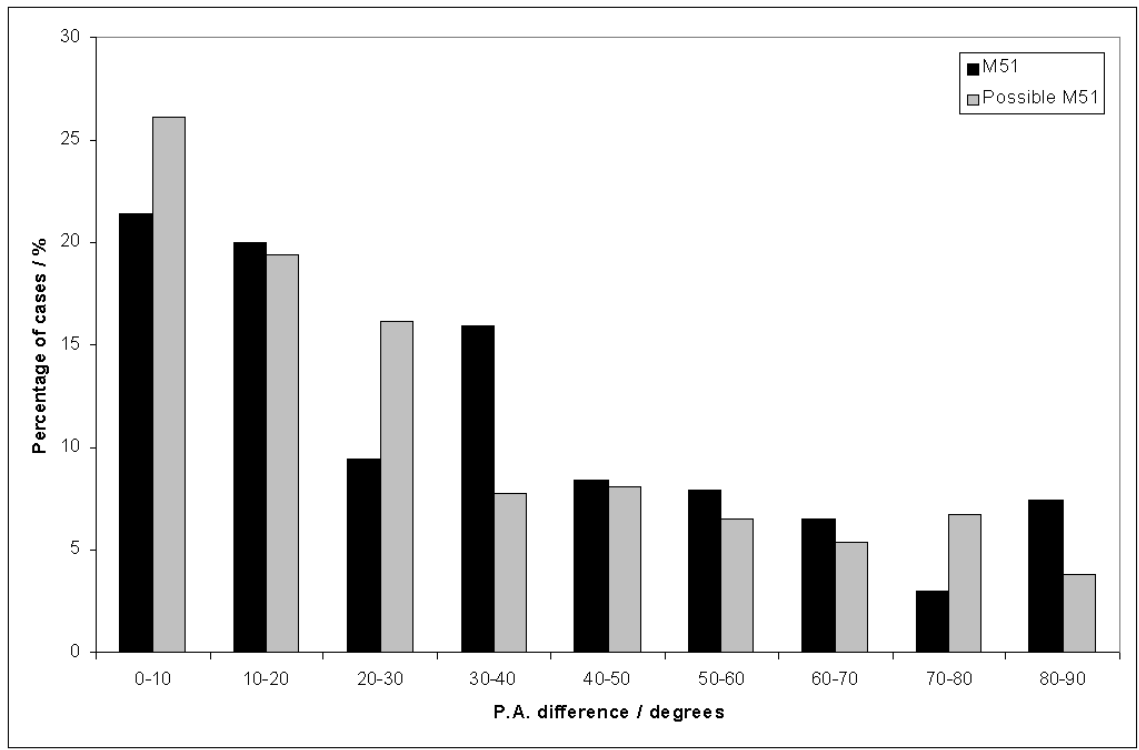

Figure 6 – Differential of companion to main galaxy major axis position angles.

### 4.8 Redshift Statistics

One significant finding during the compilation of this catalog is that whereas most of the main spirals (207/232, 89%) in the M51 associations have measured redshifts, only 41 (18%) of the companion galaxies have a published redshift. This problem was also noted in the VVII catalog. Given the potential value of these interacting pairs for the study of gravitational dynamics and galactic evolution, a systematic spectroscopic study of these systems would seem warranted. It should be noted that 17 of the 41 pairs for which the companion has a measured redshift are "discordant redshift" pairs in which the companion has a redshift at least 1000 km s$^{-1}$ larger than the main spiral galaxy. While chance projections of background galaxies may occur in a sample such as this, we have kept the discordant redshift examples in the catalog because (1) they meet the morphological criteria used to compile the catalog and (2) the extreme incompleteness of the companion redshifts in the catalog and high percentage of discordant redshift pairs suggests that a significant number of associations in the catalog for which there is currently no spectroscopy on the companions may also be discordant redshift systems. If it is assumed that the companions without measured redshifts have the same fraction of discordant redshifts, then an additional 79 discordant redshift systems would be expected when the redshift sampling of the catalog is complete. Therefore keeping the 17 discordant redshift examples in the catalog will allow for better constraints on the statistics of possible chance projections.

### 5. Conclusion

The catalog of M51 systems presented here may provide an opportunity to test for the effects of M51 type interaction on arm structure and star formation. Enhanced star formation may be present in the arm connecting to the companion in some cases. For example case #187 (NGC 6372) in the catalog shows an enhanced surface brightness of the arm with the companion relative to the opposing arm when viewed in the DSS B band. This effect is significantly reduced in the R and IR bands, which may indicate the presence of increased star formation from HII regions in the arm. This possibility would require more detailed analysis and spectroscopy and any statistically relevant study regarding the impact of M51 type interaction on star formation will require a detailed study of the examples in the catalog.

Often, bright, condensed objects appear at the ends of arms among the possible M51 candidates in Table 2. These objects have the appearance of foreground stars in chance projection with the spiral arms or HII regions within the spiral arms. Follow-up spectroscopy will be necessary to determine the natures of these objects. Examples of this phenomenon are flagged as stellar or HII in Table 2 with reasons 3 and 2 respectively in column 12 of Table 2. Prominent examples of this phenomenon include table 2 cases #346, #351, #435, and #446.

In a number of pairings, the companion galaxy is not only associated with the end of a spiral arm, but also has an apparent tidal stream pointing to an adjacent arm or to the disk of the main galaxy (eg. Table 1 cases: #16, #95, #102, #54, #189, #226, and Table 2 case #269). These features may provide additional evidence for gravitational interaction.

In this paper we have presented a catalog of M51 type associations identified based upon the examination of images in visible light and infrared – primarily survey images from the IRSA archives. We note that our catalog has surprisingly low number of early spirals and barred main galaxies. Two-armed main galaxies are clearly most common in our catalog. We show that some of the systems that were classified as M51 types in previous studies are not M51 types.

While redshift data is available for most of the main galaxies in the M51 catalog, only a small fraction of the companions have redshift data in the literature. Additional spectroscopy will therefore be needed for most of the catalog if the catalog members are to be used for tests of the effect of M51 type interactions on star formation and galaxy dynamics. For some of the cases in Table 2, additional images are unavailable and the survey images were not sufficiently resolved to allow an unambiguous acceptance or rejection of a galaxy pair. Imaging at higher resolutions may enable us to either reject some of the candidates currently populating Table 2 or accept them into the data set as M51 type associations. In the case of galaxy associations that are not unambiguously interacting gravitationally in visible-light images, follow-up observations in other bands (e.g. 21 cm emission) could identify systems that are of M51 type.

## Acknowledgements

This research has made use of the NASA/IPAC Infrared Space Archive, which is operated by the Jet-Propulsion laboratory, California Institute of Technology, under contract with the National Aeronautics and Space Administration. This research has made use of the NASA/IPAC Extragalactic Database (NED) which is operated by the Jet Propulsion Laboratory, California Institute of Technology, under contract with the National Aeronautics and Space Administration. This research has made use of the HyperLeda database (http://leda.univ-lyon1.fr). This research used the facilities of the Canadian Astronomy Data Centre operated by the National Research Council of Canada with the support of the Canadian Space Agency. This research has made use of the SIMBAD database,operated at CDS, Strasbourg, France. This research has made use of the Digitized Sky Survey. The Digitized Sky Surveys were produced at the Space Telescope Science Institute under U.S. Government grant NAG W-2166. The images of these surveys are based on photographic data obtained using the Oschin Schmidt Telescope on Palomar Mountain and the UK Schmidt Telescope. This research has made use of The NGC/IC Project website (www.ngcic.com).

This research has also made use of the Sloan Digital Sky Survey. Funding for the Sloan Digital Sky Survey (SDSS) and SDSS-II has been provided by the Alfred P. Sloan Foundation, the Participating Institutions, the National Science Foundation, the U.S. Department of Energy, the National Aeronautics and Space Administration, the Japanese Monbukagakusho, and the Max Planck Society, and the Higher Education Funding Council for England. The SDSS Web site is http://www.sdss.org/.

The SDSS is managed by the Astrophysical Research Consortium (ARC) for the Participating Institutions. The Participating Institutions are the American Museum of Natural History, Astrophysical Institute Potsdam, University of Basel, University of Cambridge, Case Western Reserve University, The University of Chicago, Drexel University, Fermilab, the Institute for Advanced Study, the Japan Participation Group, The Johns Hopkins University, the Joint Institute for Nuclear Astrophysics, the Kavli Institute for Particle Astrophysics and Cosmology, the Korean Scientist Group, the Chinese Academy of Sciences (LAMOST), Los Alamos National Laboratory, the Max-Planck-Institute for Astronomy (MPIA), the Max-Planck-Institute for Astrophysics (MPA), New Mexico State University, Ohio State University, University of Pittsburgh, University of Portsmouth, Princeton University, the United States Naval Observatory, and the University of Washington.

The authors thank Mike Petersen for his generosity in building and hosting the M51 type galaxy association website located at www.jorcat.com and for offering suggestions relating to this project. The size of the image database and the tables alone prohibit print publication of the entire project, and we are grateful for his willingness to make these materials available on-line for ready reference. We also wish to thank Tom Thomsen for several important suggestions regarding the presentation of this catalog.

The authors would also like to thank an anonymous referee for excellent suggestions for improving our work remarkably.